\documentclass{article}

\usepackage{PRIMEarxiv}

\usepackage[utf8]{inputenc} 
\usepackage[T1]{fontenc}    
\usepackage{hyperref}       
\usepackage{url}            
\usepackage{booktabs}       
\usepackage{amsfonts}       
\usepackage{nicefrac}       
\usepackage{microtype}      
\usepackage{lipsum}
\usepackage{fancyhdr}       
\usepackage{graphicx}       
\graphicspath{{media/}}     

\usepackage{algorithmic}
\usepackage{graphicx}
\usepackage{textcomp}
\usepackage{xcolor}
\def\BibTeX{{\rm B\kern-.05em{\sc i\kern-.025em b}\kern-.08em
    T\kern-.1667em\lower.7ex\hbox{E}\kern-.125emX}}

\usepackage{enumerate}
\usepackage{xspace}
\usepackage{multirow}
\usepackage{subfigure}
\newcommand{\fullname} {{\textsc{Trace2Vec}}\xspace}

\newcommand{\hide}[1]{}
\usepackage{amsmath}

\title{Detecting Complex Multi-step Attacks with Explainable Graph Neural Network
\hide{\thanks{\textit{\underline{Citation}}: 
\textbf{Authors. Title. Pages.... DOI:000000/11111.}} 
}
}

\author{
  Wei Liu \\
  Department of Computer Science and Technology\\Xiamen University\\
  NARI Group Corporation/\\
  State Grid Electric Power Research Institute\\
  Nanjing, China\\
 \texttt{liuwei5@stu.xmu.edu.cn} \\
   \And
  Peng Gao* \\
  School of Cyber Science and Engineering \\
  Nanjing University of Science and Technology \\
    Nanjing, China\\
  \texttt{gao.itslab@gmail.com} \\
  \AND
  Haotian Zhang, Ke Li, Weiyong Yang, Xingshen Wei\\
  NARI Group Corporation/\\
  State Grid Electric Power Research Institute\\
  Nanjing, China\\
  \texttt{\{zhanghaotian1, like3,yangweiyong,weixingshen\}@sgepri.sgcc.com.cn} \\
  \And
  Jiwu Shu \\
Department of Computer Science and Technology\\Xiamen University\\
    Xiamen, China \\
  \texttt{jwshu@xmu.edu.cn} \\
}

\begin{document}
\maketitle

\begin{abstract}
Complex multi-step attacks have caused significant damage to numerous critical infrastructures. To detect such attacks, graph neural network based methods have shown promising results by modeling the system's events as a graph. However, existing methods still face several challenges when deployed in practice. First, there is a lack of sufficient real attack data especially considering the large volume of normal data. Second, the modeling of event graphs is challenging due to their dynamic and heterogeneous nature. Third, the lack of explanation in learning models 
undermines the trustworthiness of such methods in production environments. 
To address the above challenges, in this paper, we propose an attack detection method, \fullname.  
The approach first designs an erosion function to augment rare attack samples, and integrates them into the event graphs.
Next, it models the event graphs via
a continuous-time dynamic heterogeneous graph neural network. Finally, it employs the Monte Carlo tree search algorithm to identify events with greater contributions to the attack, thus enhancing the explainability of the detection result. We have implemented a prototype for \fullname, and the experimental evaluations demonstrate its superior detection and explanation performance compared to existing methods.
\end{abstract}

\keywords{Complex multi-step attack \and Anomaly detection \and Dynamic heterogeneous graph \and Attack sample construction \and Explainability}

\section{Introduction}
Nowadays, complex multi-step attacks (CMAs) have become one of the main threats against numerous critical information infrastructure~\cite{navarro2018systematic}. One typical example is the WannaCry ransomware attack~\cite{chen2017automated} that broke out in May 2017, which was reported to infect more than 230,000 computers in over 150 countries within one day. 
Such attacks are usually composed of multiple single-step attacks, and employ a blend of sophisticated techniques, such as supply chain attack~\cite{martinez2021software} and social engineering~\cite{krombholz2015advanced}, to exploit unknown vulnerabilities in a system.

The complexity of CMAs presents substantial challenges for traditional detection methods.
To this end, extensive data-driven research efforts have been conducted in both academia and industry~\cite{Hossain2017SLEUTH,Milajerdi2019POIROT,Rossi2020TGN,Han2020Unicorn,Abdulellah2021ATLAS,Zhu2021APTSHIELD,Yang2022RShield,Ding2023AIRTAG}. 
Among them, graph neural network (GNN)-based methods~\cite{Abdulellah2021ATLAS,Yang2022RShield,Wei2021Deephunter,Liu2019Log2vec,Yang2022Advanced}, which treat the system events as an event graph, have shown promising results for detecting CMAs.
However, the existing methods still suffer from several limitations, especially when deploying them in production environments. 
\begin{itemize}
    \item First, in real systems, attack behaviors are extremely sparser than normal behaviors. This may cause learning-based methods (e.g., GNNs) to be less effective as they have to be trained on very imbalanced data. 
    \item Second, CMA detection involves temporal and heterogeneous data, where attack patterns may span over a long time period, making it difficult for GNNs to capture the attack behavior. 
    \item Third, the ``black-box'' nature of GNN models raises the explainability issue, which is crucial for decision-making in critical cybersecurity applications, and the lack of which may lead to untrustworthy detection results.
\end{itemize}

To address the above limitations, we propose a CMA detection method, namely \fullname, using explainable deep spatio-temporal graph neural networks. 
Specifically, \fullname first leverages prior attack knowledge to generate strong negative samples (i.e., attack behaviors) to mitigate the label imbalance problem. 
Subsequently, \fullname trains an attack detection model by utilizing an existing continuous-time dynamic heterogeneous graph (CDHGN) modeling technique~\cite{Gao2022Detecting} on the event graph constructed from the audit logs, and further improves it with a fine-tuning stage.
Finally, \fullname combines the Monte Carlo tree search algorithm to identify abnormal and normal events with higher contributions to the attack behavior, enhancing model explainability. 
We condunct experiments on the widely-used synthetic dataset CERT~\cite{Lindauer2020CERT} and 
the real-world APT attack dataset ATLAS~\cite{Abdulellah2021ATLAS}. The results show that \fullname has better detection and explanation performance than current state-of-the-art methods.

The main contributions of this paper include:
\begin{itemize}
    \item A CMA detection framework addressing three key challenges of data imbalance, graph modeling, and prediction explainability, which are crucial for the  deployment in production environments.
    \item Provision of explainable results for detecting CMAs. To the best of our knowledge, this represents the first research effort to integrate GNN explainability technique for CMA detection.
    \item The experimental evaluations demonstrate that the proposed method has improved the detection accuracy by up to 13.0\% compared to other methods, while enhancing the explainability of the detection results by up to 20.1\%.
\end{itemize}

The structure of the paper is as follows. Section 2 covers related work. Section 3 presents the overview of the proposed method, and the details are provided in Section 4 and 5. Section 6 presents the experimental results. Section 7 concludes the paper and suggests future research directions.

\section{Related Work}
\subsection{Attack Detection and Investigation}
Many studies utilize GNNs to identify complex multi-step attacks. These methods construct an event graph from logs (e.g., a user logons a web service is represented as an edge in the graph), and excel in capturing nuanced network behavior patterns among network entities. 
For instance, Log2vec~\cite{Liu2019Log2vec} leveraged heuristic rules to establish connections between log records, thereby constructing a network behavior graph. Subsequently, graph embeddings were generated to discern between malicious and benign activities using clustering algorithms applied to node embeddings. 
To improve the model's generalization and achieve end-to-end detection, RShield~\cite{Yang2022RShield} and CTHGN~\cite{Gao2022Detecting} introduced methods aimed at detecting CMAs, leveraging homo/heterogeneous temporal graph network models. These frameworks integrate attention and memory mechanisms into dynamic graph information propagation, facilitating anomaly detection in continuous-time dynamic graphs. Nonetheless, owing to the constraints associated with repeat sampling, their ability to capture contextual information pertaining to network entity behaviors is limited.

Attack investigation extracts attack information from audit logs, network traffic, or other data to reconstruct the entire attack process. Existing works, such as SLEUTH~\cite{Hossain2017SLEUTH}, focus on generating and analyzing provenance graphs from audit logs and using tag-based techniques to detect attacks. It also provides a visual graph of attack steps for traceability. APTSHIELD~\cite{Zhu2021APTSHIELD} uses ATT\&CK to construct an information aggregation framework and APT detection rules from system data flow, enabling real-time alarms and response. However, its reliance on prior knowledge and expert rules poses challenges in detecting unknown attacks.
ATLAS~\cite{Abdulellah2021ATLAS} is a sequence-based model combining causal analysis for traceability. However, it suffers from a lot of manual labeling, coarse granularity, and limited explainability. AIRTAG~\cite{Ding2023AIRTAG} addresses these issues using an unsupervised method based on BERT, integrating time information while without spatial details. Although the provenance graph-based methods simplify the process of attack investigation, the graphs are still large scale and hard to understand.

\subsection{Graph Explainer}
Few methods currently exist for explaining temporal graph models, with most related works focusing on static GNNs. Typically, existing methods analyze the changes in the output of the model by perturbing the inputs of the GNN and identify graph components that significantly impact the prediction results~\cite{Yuan2020Xgnn}. Existing methods generally fall into two categories: learning-based methods and search-based methods.

Learning-based methods like GNNExplainer~\cite{Ying2019Gnnexplainer}, PGExplainer~\cite{Luo2020Parameterized}, 
Reinforced explainer~\cite{Shan2021Reinforcement}, 
and PGM-Explainer~\cite{Vu2020Pgm-explainer} often use neural networks to identify key nodes/edges based on the representation generated by trained GNN models. Specifically, GNNExplainer identifies a compact subgraph structure and a small subset of node features for explanations. PGM-Explainer identifies crucial graph components, demonstrating dependencies through conditional probabilities. Although effective in identifying influential nodes, edges, and subgraphs on predictions, learning-based methods struggle to capture the correlation between spatio-temporal dependencies in the graph and prediction results.

Search-based methods, such as SubgraphX~\cite{Yuan2021OnExplainability} and Causal-Interpret~\cite{Wang2020Causal}, use searching algorithms with a score function to find important input subsets. For example, SubgraphX, employing Monte Carlo tree search (MCTS), explores subgraphs effectively for explanations. 
Other methods are also proposed. For example, He et al.~\cite{He2022explainer} use a probability graph model for an explainer framework. Xia et al. introduce TGNNExplainer~\cite{Xia2023Explaining}, employing MCTS to find a subset of historical events for predictions in a temporal graph. However, TGNNExplainer struggles to capture deep correlations in computer network entity interactions over a long time span, impacting explanation accuracy for CMAs. Additionally, its random masking of edges is limited to predicting future edge occurrences, not detecting attack behaviors.

\section{Approach Overview}
\begin{figure}[htb]
    \centering
    \includegraphics[width=0.98\textwidth,height=0.38\textwidth]{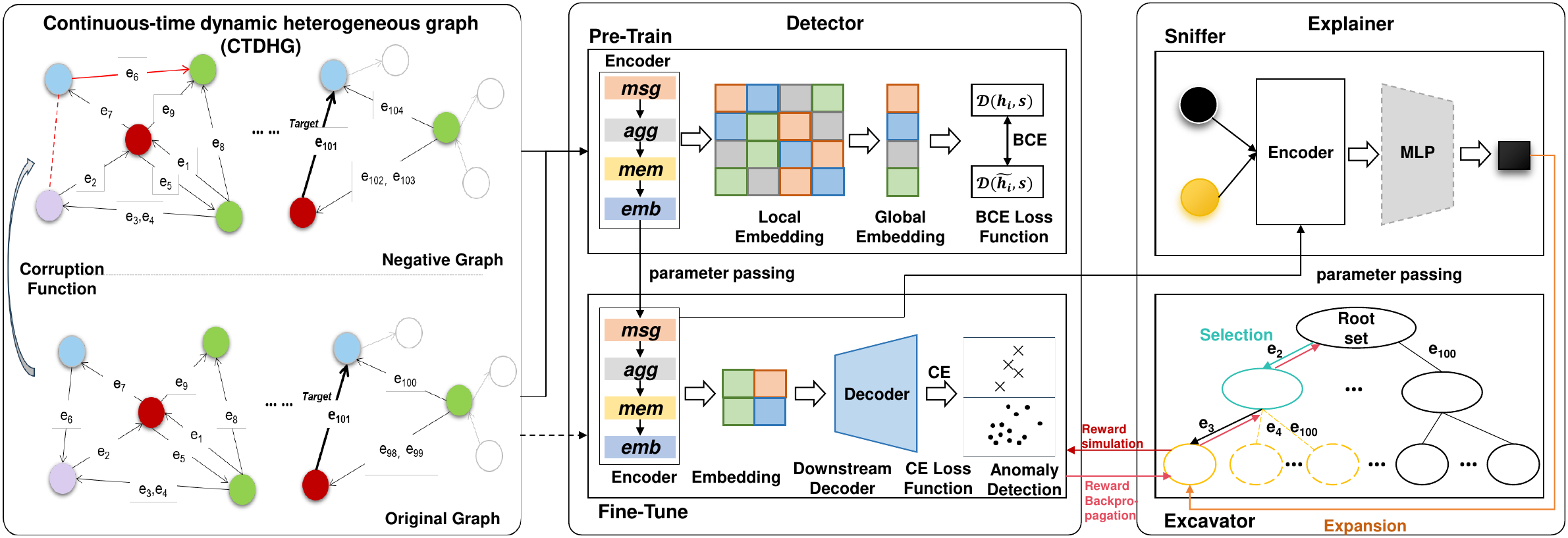}
    \caption{Overview of the \fullname framework. In CTDHG, different colors of nodes represent different node types. $e_i$ denotes an event occurring at time $i$ (directed edge). Red dashed and solid lines represent deleted and newly generated edges via the erosion function, respectively. Black solid directed edges represent the target events. For simplicity, gray nodes and edges denote omitted other nodes and edges.}
    \label{fig:overview}
\end{figure}

The overview of \fullname is illustrated in Fig.\ref{fig:overview}. Specifically, \fullname is primarily composed of a {\em detector} and an {\em explainer}.

\begin{itemize}
    \item The detector first designs an erosion function to augment the extremely rare attack events (negative samples). It then pre-trains a graph neural network based on the augmented event graph. Next, the detector is fine-tuned using a small number of real attack samples, to make it more sensitive towards the attack behaviors. 
    \item The explainer consists of a sniffer and an excavator. The sniffer leverages the output of the detector and further trains a Multilayer Perceptron (MLP) to compute the relevance score of an input event to the detected attack behavior. The excavator applies Monte Carlo Tree Search~(MCTS) to search for a subgraph of events that could explain the current attack behavior based on the relevance scores. 
\end{itemize}

Here, we provide the key symbols and denotations used throughout the paper.
We let $S = \left \{ E_1, E_2, \cdots \right \}$ denote a sequence of events. Our method aims to train a GNN model to detect the anomalous event $E_k$ and explain the result by identifying a subset of events $Exp_k$ for explanation.
Each event $E_i = \left \{ v^{p}_{src_i}, v^{q}_{dst_i}, e_{i}, t_i \right \}$ comprises two nodes $v^{p}_{src_i}$ (source node with type $p$), $v^{q}_{dst_i}$ (destination node with type $q$), an edge $e_{i}$ between two nodes (id and attribute), and timestamp $t_i$. 
Here, the type of a node refers to the category or class that a node belongs to within the graph.
The sequence of events also can be represented as a continuous-time dynamic heterogeneous graph (CTDHG). 
Similar to the setting in~\cite{Xia2023Explaining}, we use $g^{i}$ to denote the graph constructed from the events $\left \{ E_1, E_2, \cdots, E_{i-1} \right \}$
excluding $E_i$. In this context, our approach could be formulated by the following steps:
\begin{enumerate}
    \item The detector takes $g^{i}$ as input, and learns the embedding representation $Z_i = \left \{ z_{1}, z_{2}, \cdots \right \}$, where $z_{j}$ denotes the current embedding of node $n_{j}$ (i.e., $v^{p}_{src_i}$ or $v^{q}_{dst_i}$) at timestamp $t_i$. 
    \item The decoder then takes the node embeddings as input to detect the attack (i.e., anomalous event) $E_k$. 
    \item To trace the attack, the explainer takes node embeddings and anomalous event $E_k$ as input, and utilizes a search-based method to identify the event subset $Exp_k$ for explanation. 
\end{enumerate}

\section{Detection Module}
The detection module of \fullname consists of three stages: negative sample generation, pre-training, and fine-tuning. 
The first stage is to construct the samples of attack pattern graph based on cyber threat intelligence (CTI) reports, the second stage is to pre-train the encoder for the CMA detector via contrastive learning, and the last stage is to fine-tune the encoder and decoder on a small set of realistic attack samples.

\subsection{Negative Sample Generation Stage}
Real-world datasets often exhibit severe class imbalance issue that makes it difficult to utilized in a supervised learning model. To address this, we propose an improved negative sample generation approach that eliminates the need for pre-defined rules 
used in ~\cite{Gao2023Deep}. The method involves constructing provenance and attack pattern graphs, aligning these graphs, and augmenting the overall graph.

\textit{Provenance Graph Construction.} Normal system behaviors are modeled as a provenance graph labeled and directed through logs. Entities such as files and processes serve as nodes, 
and relationships between entities are represented by directed edges. Both nodes and edges have associated features. We employ an existing NLP-based approach~\cite{Hossain2017SLEUTH} to parse system logs and construct the provenance graph.

\textit{Attack Pattern Graph Construction.} We resort to the CTI reports to construct the attack pattern graphs~\cite{Milajerdi2019POIROT}. 
Specifically, we use the multi-class classifier introduced in \cite{zhu2018chainsmith} to extract IOCs. 
Besides, we also develop a rule-based method to identify domain-specific characters such as CVE numbers and IP addresses. 
After that, NLP parsing is employed to extract the dependencies between IoC entities and construct attack pattern graphs similar to provenance graphs. 

\textit{Graph Alignment.} We next search the provenance graph for subgraphs which are similar to the attack pattern graph.  
This involves matching nodes based on their features in the provenance graph and using a depth-first search strategy to find reachable node sets and aligned paths~\cite{Milajerdi2019POIROT}. 

\textit{Graph Augmentation.} Finally, we augment the original provenance graph by incorporating the attack pattern graph. 
In order to embed prior knowledge of CTI into normal system behaviours, we adopt different graph augmentations to construct attack samples based on the attack pattern graph and the aligned subgraph. Specifically, the graph augmentations include node perturbation, edge perturbation, feature masking, and random walks. Node perturbation removes or introduces nodes from/to the aligned subgraph, while edge perturbation does the same for edges. Feature masking conceals features of randomly chosen nodes in the aligned subgraph, and random walks generate new subgraphs from related nodes/edges in the provenance graph. These techniques aid in constructing realistic attack samples. 

\subsection{Pre-training Stage}
After generating negative samples, we pre-train the encoder for the CMA detector using contrastive learning upon the continuous-time dynamic heterogeneous graph (CDHGN)~\cite{Gao2022Detecting}. 
We use contrastive learning as it can output more robust node embeddings. We also incorporate the transformer-XL attention mechanism to capture ultra-long context dependencies and employ a neural temporal point process model for contextual causality between events in continuous time.

We denote the positive sample (provenance graph representing normal system behaviors) as $(X, A)$, where $X$ is the node feature matrix and $A$ is the adjacency matrix. The corresponding negative samples generated based on $(X, A)$ are denoted as $(\tilde{X}, \tilde{A})$.
The CDHGN encoder uses message passing, aggregation, memory update, and embedding generation to obtain the local feature $H = \varepsilon (X, A) = ( { h_1, h_2, \dots, h_n } )$,
where $\varepsilon$ represents the encoder and $h_i$ represents the embedding for node $i$. 
Then, CDHGN uses a readout function \textit{R} to aggregate all obtained local features as the global feature. Referring to DGI~\cite{Velickovic2019DGI}, our method averages the local features of all nodes as $\textit{R}(H) = \sigma(\frac{1}{n} \sum^{n}_{i=1} h_i)$.

Based on the global feature $s = \textit{R}(H)$, we aim to maximize the mutual information between the representations of positive examples and the overall graph. 
Hence, the noise-contrastive type objective with a standard binary cross-entropy (BCE) loss is used as the loss function in equation~\eqref{eq:nceloss}:
\begin{equation}
\label{eq:nceloss}
\begin{split}
   BCE Loss = \frac{1}{N+M}\left( \sum_{i=1}^{N}E_{(X,A)}\left [ \log D \left( h_i, s \right ) \right ]  \right. \\
   \left. + \sum_{j=1}^{M}E_{(\tilde{X},\tilde{A})}\left [ \left ( 1-\log D \left( \tilde{h_j}, s \right ) \right ) \right ] \right ) 
\end{split}
\end{equation}
where $D$ denotes a discriminator function; N and M denote the number of nodes in positive example and negative example,  respectively. 

\subsection{Fine-tuning Stage} \label{subsec:fine-tuning}
After pre-training, the encoder parameters are passed to the fine-tuning stage, where both the encoder and decoder adapt to a limited set of realistic attack samples. 
According to its type, the embedding of an edge is obtained by splicing the embedding of the nodes on both sides of the edge.
Then, the fully connected layer maps the embedding into the probability that a connected edge belongs to an attack event. 
The cross-entropy loss function in equation~\eqref{eq:celoss} is used as the fine-tuning goal, where $y_x$ denotes the ground truth whether edge $x$ is an attack event, and $p_x$ denotes the predicted probability of $x$ belonging to an attack event. 
\begin{equation}
\label{eq:celoss}
    CE Loss = -\sum_{x} \left[ y_x\log(p_x) + (1-y_x) \log (1-p_x) \right]
\end{equation}

\section{Explanation Module}
In this section, we explore how to identify the explainable subgraph $Exp_k$ corresponding to the detected anomalous event $E_k$.
Any events preceding $E_k$ could potentially serve as part of the explanation. 
To efficiently identify key steps related to the attack behavior, we enhance the TGNN (TGNNExplainer) method~\cite{Xia2023Explaining} and propose the ``Sniffer-Excavator'' framework.
As illustrated in the right part of Fig.\ref{fig:overview}, the sniffer is trained to compute the relevance between the anomalous event and the preceding events. The excavator, guided by the sniffer, utilizes the Monte Carlo tree search to output the final subgraph $Exp_k$. 

\subsection{Sniffer}
As mentioned above, methods such as~\cite{He2022explainer,Xia2023Explaining} employ random masking to generate event embeddings, inferring relevance scores of events related to the target. However, such an approach is not suitable for measuring the correlation of the attack event with others, because the assumption that nonexistent edges indicate attack is invalid. Therefore, we build a ``threat'' sniffer to compute the relevance score of an event $E_j$ to the attack event $E_k$. The sniffer denoted as $H_\theta(E_j,E_k)$, inherits the encoder from the fine-tuning stage of the detector (Section \ref{subsec:fine-tuning}), and combines it with MLP layers.
These scores are utilized to optimize the MCTS during the excavator node expansion process. 

The input and the training process are described below. Let the target event be $E_k = \left \{ v^{p}_{src_k}, v^{q}_{dst_k}, e_{k}, t_k \right \}$  and each candidate event $E_j = \left \{ v^{p}_{src_j}, v^{q}_{dst_j}, e_{j}, t_j \right \}$, the input of the sniffer is denoted as $Z_{E_k,E_j} = \left[ vec(E_k), vec(E_j)\right]^T$ where $vec(\cdot)$ is a function that transforms raw event data into a vector. 
Here, we let $vec(E_i)$ be $\left[X_{v^{p}_{src_i}}, X_{v^{q}_{dst_i}}, Time(t_i), e_k\right]$, in which the $X$ is the node feature matrix, and $Time(\cdot)$ is a learnable function that embeds real-valued timestamps into a vector.
We adopt the relative dynamic time-encoding function~\cite{Gao2022Detecting} 
to calculate the time embedding over an interval: $
\operatorname{RTE}(\Delta T)=\operatorname{TimeLinear}(\operatorname{RT}(\Delta T))$. Here, $\Delta T$ denotes 
the position, the $TimeLinear$ is a learnable linear projection function, and the $\operatorname{RT}(\Delta T)$ is the general term for 
 $ \operatorname{RT}(\Delta T, 2 \operatorname{dim})=\sin \left(\frac{\Delta T}{10000^{(2 \operatorname{dim} / d)}}\right)$
and $\operatorname{RT}(\Delta T, 2 \operatorname{dim}+1)=\cos \left(\frac{\Delta T}{10000^{((2 \operatorname{dim}+1) / d)}}\right)$. The dim is the dimension denoting each
dimension of the positional encoding corresponds to a sinusoid.
The relative-time embedding enhances the feature representation of both the source and target nodes, facilitating the capture of dynamic information pertaining to the temporal relationship between these nodes. 

We input the detected attack event and each candidate event (i.e., all events before the attack event) into $H_\theta$, and get the output which is the corresponding  relevance score between them.

\subsection{Excavator}
In the excavator module, we use the MCTS algorithm to pinpoint events closely related to the attack. First, we initialize the root node with a set of candidate events. Then, multiple round simulations are conducted to expand the nodes in the search tree, where each node represents an event subset. Each round consists of four key steps: 
\begin{itemize}
    \item \textbf{Selection.} Starting from the root (node set), select child nodes iteratively until reaching a leaf node based on a certain strategy.
    \item \textbf{Expansion.} Expand the selected leaf node to generate new child nodes by removing unimportant events according to the detector suggestion.
    \item \textbf{Simulation.} Simulate the reward for the new child node using the temporal graph model of the detector. 
    \item \textbf{Backpropagation.} Propagate the reward from the leaf node upward, updating reward of each node along the path to the root. 
\end{itemize}

Finally, the node set that achieves the highest reward is considered the explanation result. The detailed mechanism is explained as follows.

\textit{Initialization.} The root node is a set includes candidate events' edge satisfying spatial and temporal conditions. For instance, suppose the target is explaining the edge of attack event $e_{100}$ in Fig.\ref{fig:overview}. For tracing attack steps, the root node set of events seen by the encoder is $\{e_1, e_2, ..., e_{99}\}$.

\textit{Node Selection.} At each MCTS search node $N_i$, for a removal action $e^*$ (removing $e^*$ from $N_i$), we use the Upper Confidence Bound applied to Trees (UCT)~\cite{Kocsis2006Bandit} to balance exploitation and exploration during node selection. Supposing at node $N_i$ , the criterion for the action is: 
\begin{equation}
\label{eq:ucb}
    e^*=\underset{e_j \in \mathcal{C}\left(N_i\right)}{\arg \max }\left(\frac{c\left(N_i, e_j\right)}{n\left(N_i, e_j\right)}+\lambda \frac{\sqrt{\sum_{e_l \in \mathcal{C}\left(N_i\right)} n\left(N_i, e_l\right)}}{1+n\left(N_i, e_j\right)}\right)
\end{equation}
where $C(N_i)$ represents the set of events already expanded under node $N_i$, $n(N_i,e_j)$ is the number of times the removal action $e_j$ has been chosen at node $N_i$ in previous simulations, and $c(N_i,e_j)$ represents the cumulative reward for choosing $e_j$ at node $N_i$. The formula balances exploitation (left part) by choosing nodes with higher average rewards and exploration (right part) by selecting nodes that have been simulated fewer times. The algorithm moves to a child node of $N_i$ by removing the event $e^*$ from $N_i$.

\textit{Node Expansion.} To optimize the search space and improve node selection quality, we adopt a strategy to expand only nodes with potential relationships with the attack event. Instead of expanding all possible child nodes at each selected node~\cite{Yuan2021OnExplainability}, we narrow down the search space by the sniffer. If the selected node $N_i$ is deemed expandable (i.e., child node), the excavator calculate potential scores:
\begin{equation}
\label{eq:expand}
    e^*=\underset{e_j \in \left(N_i - \mathcal{C}\left(N_i\right)\right)}{\arg \min } H_\theta\left(e_j, e_k\right)
\end{equation}
where $N_i$-$C(N_i)$ denotes events that have not been expanded in previous simulations. We retain the most relevant events by discarding the least important event(s) $e^*$ to expand a new node.
Leveraging the pre-trained sniffer inherited from the detector, we efficiently infer scores with minimal computational cost. The selection and expansion of nodes occur in an alternating fashion. Starting from the root node, we expand new child nodes based on the above equation~\eqref{eq:expand}. Then we select the node with the highest score from both the original and new child nodes of the root node and move to that node according to equation~\eqref{eq:ucb}. Next, we repeat the process of expansion and selection from the new node. The process ends when the current node is determined as a leaf node or when the number of nodes is less than the hyperparameter $k$.

\textit{Simulation and Backpropagation.} The reward is simulated by the detection model, with $r(N_{leaf})$ calculated for the leaf node $N_{leaf}$. During the backpropagation, all nodes $N_i$ from the root to the leaf update $n(\cdot, \cdot)$ and $c(\cdot, \cdot)$ by accumulating the reward from the leaf node and incrementing by 1. Specifically, $c(N_i, e^*) = c(N_i, e^*) + r(N_{leaf})$ and $n(N_i, e^*) = n(N_i, e^*) + 1$.

\section{Experimental Evaluation}
In this section, we present the experiment results. The experiments are designed to answer the following questions.
\begin{enumerate}[RQ1:]
    \item Can the proposed method accurately detect complex network attacks?
    \item Is it effective in explaining key steps associated with the detected attack?
    \item While MCTS improved real-time detection, it posed a challenge in tracing attack steps over long-time intervals. Can \fullname address this issue?
\end{enumerate}

\subsection{Experimental Setup}
\subsubsection{Datasets}
We use the CERT dataset (V6.2)~\cite{Lindauer2020CERT} in our experiments. It contains labeled insider threat activities, generated by sophisticated user models in a simulated organization's computer network. It comprises five log files capturing various computer-based activities (logon/logoff, HTTP, email, file operations, and external storage device usage) from 4,000 users over 516 days. For experiments, 41 users were randomly sampled, providing 1,306,644 events, with 470 manually labeled events representing five insider threat scenarios. The dataset also includes user attribute metadata with six categorical attributes, offering additional context. 
In each of the five attack scenarios, only a single malicious user performs a series of attack steps.  
\hide{The original data stores logs in separate files, necessitating integration for feature extraction. Attributes including role, project, functional unit, department, team, supervisor, email addresses, specific behaviors, and timestamps, serve as data features. Statistical features include indicators like device usage outside working hours, job departure within two months, and access to suspicious websites.}

We also utilize public datasets containing real-world APT attack scenarios sourced from ATLAS (M2 and M5)~\cite{Abdulellah2021ATLAS}, encompassing system, DNS, and Firefox logs. The M2 dataset corresponds to a \textit{phishing} attack directed at web browsers, while the M5 dataset represents a \textit{pony campaign} targeting Microsoft Office. The sizes of M2 and M5 are 671.112 MB and 726.113 MB, respectively. The M2 dataset contains 13.68\% malicious behavior, while the M5 dataset contains 5.33\% malicious behavior. Both datasets encompass attacks conducted across multiple victim hosts.
Initially, the attacker replaced a benign webpage within the victim's system with a malicious one on the first host. Subsequently, on the second host connected to the first, the victim accessed the malicious webpage, resulting in compromise.
Each attack targeting multiple hosts results in the collection of log files obtained from the respective victim host.

\subsubsection{Baselines}
For detection accuracy, we compare with three baseline methods, i.e., Log2vec$\backslash$Log2vec++~\cite{Liu2019Log2vec}, TGNN (TGNNExplainer)~\cite{Xia2023Explaining}, and CDHGN~\cite{Gao2022Detecting}. Log2vec employs a rule-based heuristic approach for constructing a heterogeneous graph and extracting features using improved random walks. TGNN directly employs
continuous-time dynamic graph network (CDGN), developed by TGN\cite{Rossi2020TGN}, to predict the existence of edges in the event graph. Building upon this, we adjust the objective function of TGNN from "predicting the existence of edges" to "predicting whether edges are anomalous," and employ resampling techniques, similar to RShield\cite{Yang2022RShield} and CDHGN\cite{Gao2022Detecting}, to balance the number of positive and negative samples. CDHGN integrates heterogeneous attention and memory into dynamic graph information propagation for anomaly detection in heterogeneous event graphs. 

For explainability, we include baselines of 
GNNExplainer~\cite{Ying2019Gnnexplainer}, PGExplainer~\cite{Luo2020Parameterized}, and TGNN (TGNNExplainer)~\cite{Xia2023Explaining}. GNNExplainer generates masks to capture key semantic information, learning soft masks for edges and nodes through optimization. PGExplainer utilizes a generative probability model to learn concise latent structures, modeling latent structure as edge distribution and providing model-level explanations based on the GNN model. TGNNExplainer 
employs MCTS to identify a subset of historical events for predictions in a temporal graph.

\subsubsection{Metrics}
For attack detection, the Area Under the Curve (AUC) is utilized as a performance metric. AUC measures the accuracy of the classifier in determining whether positive samples are ranked higher than negative samples. It ranges from 0 to 1, with a higher value indicating better detection performance. 

To evaluate the explainability, the \textit{fidelity} and \textit{sparsity} are widely considered~\cite{Xia2023Explaining,yuan2022explainability}. 
Here, the fidelity reflects the model's accuracy in faithfully preserving the attack related behaviors from the input data. 
Let $N_e$ be the number of edges, $\mathbb{I}$ be the indicator function, $f(\cdot)$ be the neural network model outputting the graph embedding, $t_i$ be the time point, $R^i$ be the subgraph sampled at time $t_i$, and $G^i$ be the original graph at time $t_i$.
The fidelity definition is shown as 
\hide{
\begin{eqnarray}
fidelity(f(R^i)[t_i],f(G^i)[t_i] =~~~~~~~~~~~~~~~~~ \nonumber\\ 
\frac{1}{N_e} \sum_{i=1}^{N_e}\Big [\mathbb{I}\left(Y_i=1\right)\left(f(R^i)\left[t_i\right] -f(G^i)\left[t_i\right]\right)\\
+\mathbb{I}\left(Y_i=0\right)\left(f(G^i)\left[t_i\right]-f(R^i)\left[t_i\right] \right) \Big ].~~~~~~ \nonumber
\end{eqnarray}
}

\begin{eqnarray}
fidelity(f(R^i)[t_i],f(G^i)[t_i] =  
\frac{1}{N_e} \sum_{i=1}^{N_e}\Big [\mathbb{I}\left(Y_i=1\right)\left(f(R^i)\left[t_i\right] -f(G^i)\left[t_i\right]\right)\\  ~~~~~~ \nonumber
+\mathbb{I}\left(Y_i=0\right)\left(f(G^i)\left[t_i\right]-f(R^i)\left[t_i\right] \right) \Big ]
\end{eqnarray}

In addition to fidelity, effective explanations should be concise, focusing solely on capturing the most pertinent input features while disregarding any irrelevant ones. This characteristic is quantified by the metric sparsity, which specifically evaluates the proportion of features deemed significant by explanation methods~\cite{pope2019explainability}.
Here, sparsity is defined as $|R^i|/|G^i|$, denoting the proportion of zeros in the model parameters or activation values. A sparse model exhibits a scarcity of parameters or activation values that are deemed crucial for the final prediction between normal and abnormal behaviors.
Higher fidelity and increased sparsity signify a better explanation. We plot the fidelity-sparsity curve and calculate the area under the curve (AUFSC) to assess performance. A larger AUFSC suggests better performance.
\hide{
As presented in equation~\eqref{eq:sparsity}, sparsity indicates the proportion of zeros in the model parameters or activation values. A sparse model has only a few parameters or activation values crucial for the final prediction or representation.
\begin{equation}
\label{eq:sparsity}
\footnotesize
\text { sparsity }=\left|R^i\right| /\left|G^i\right|
\end{equation}
}

\subsubsection{Implementations}
We implemented the \fullname prototype using Python 3.9.0 and PyTorch 1.7.0. The experiments were carried out on a PC workstation with an Intel Core i9 2.8GHz, 32GB RAM, and Windows 11 64 bit. The GPU was an Nvidia RTX3090 with 24GB VRAM. The Adam optimizer is employed with a learning rate of 0.005 and weight decay of 0.0001. The balancing coefficients for the regularization terms are set to (0.001, 0.01). Training, validation, and test sets are divided based on timestamp, considering the number of attack samples.

\subsection{Experimental Result}
\subsubsection{Effectiveness comparison}
To address RQ1, Table~\ref{tab:detect} and Table~\ref{tab:detect_atlas} present the detection outcomes of baseline models and \fullname on the CERT and ATLAS datasets, respectively. The dataset division denotes the count of malicious samples (rare negative samples) within the respective data splits. For Log2vec, our comparison focused solely on a subset of the results detailed in the paper~\cite{Liu2019Log2vec}. 

\begin{table*}[htbp]
\centering
\caption{Detection results with different divisions of malicious samples on the CERT dataset. \fullname outperforms the existing competitors.}
    \begin{tabular}{cccccccc}
        \hline
        \multicolumn{3}{c}{\bf Malicious samples division} & \multicolumn{5}{c}{\bf Method}                \\ \hline
        {\bf \#Train}     & {\bf \#Val}     & {\bf\#Test}     & {\bf Log2vec} & {\bf Log2vec++} & {\bf TGNN}    & {\bf CDHGN} &{\bf \fullname} \\ \hline
        78          & 73        & 319        &    -     &   -        & 0.9318 & 0.9514 & {\bf 0.9675}   \\
        84          & 182       & 204        &     -    &   -        & 0.9468 & 0.9675 & {\bf 0.9755}   \\
        128	        &168	   & 174         &     -    &   -        & 0.9469 & 0.9676 & {\bf 0.9724}   \\
        186	        &152	   & 132         &     -    &   -        & 0.9469 & 0.9676 & {\bf 0.9724}   \\
        234         & 132       & 104        & 0.8600  & 0.9300    & 0.9438 & 0.9620 & {\bf 0.9719}   \\ \hline
    \end{tabular}
\label{tab:detect}
\end{table*}

\begin{table*}[htbp]
\centering
\caption{Detection results with different divisions of malicious samples on the ATLAS dataset. \fullname outperforms the existing competitors.}
    \begin{tabular}{ccccccc}
        \hline
                    & \multicolumn{3}{c}{\bf Malicious samples division} & \multicolumn{3}{c}{\bf Method}   \\ \hline
                    & {\bf \#Train}     & {\bf \#Val}     & {\bf \#Test}      & {\bf TGNN}    & {\bf CDHGN} & {\bf \fullname} \\ \hline
\multirow{3}{*}{M2} & 8923       & 3569        & 5354     & 0.9101 & 0.9069 & {\bf 0.9326}   \\
                    & 10708          & 3569       & 3569        & 0.9097 & 0.9146 & {\bf 0.9395}   \\
                    & 12494         & 2676       & 2676        & 0.9108 & 0.9172 & {\bf 0.9439}   \\ \hline
\multirow{3}{*}{M5} & 8301       & 2825        & 5907     & 0.8719 & 0.8842 & {\bf 0.9498}   \\
                    &10219          & 3407       & 3407        & 0.8820 & 0.8912 & {\bf 0.9501}   \\
                    &11923         & 2555       & 2555        & 0.8829 & 0.8987 & {\bf 0.9549}   \\ \hline
    \end{tabular}
\label{tab:detect_atlas}
\end{table*}

It can observed that the proposed method outperforms the baselines across various dataset divisions on both datasets.
For example, on the CERT dataset, \fullname achieves an AUC of 0.9719 with the ratio of 234:132:104. This represents 13.0\%, 4.5\%, 3.0\%, and 1.0\% relative improvements over Log2vec, Log2vec++, TGNN, and CDHGN, respectively. On the ATLAS dataset, the maximal improvement over TGNN and CDHGN becomes 8.9\% and 7.4\%, respectively. This improvement is even larger than that on the CERT data, mainly due to the fact that the ATLAS dataset is much larger.
Among the competitors, TGNN employs a homogeneous graph model(CDGN), to predict anomalous events, achieving significant performance improvement compared to Log2vec/Log2vec++, which construct graphs based on heuristic rules. Building upon this, CDHGN further enhances performance by incorporating heterogeneous attention mechanisms for both nodes and edges in the continuous-time dynamic graph, and thus outperforms the others relatively. Compared to CDHGN, we primarily introduced a pre-training stage with a data augmentation module based on negative sampling into \fullname. The results demonstrate that our design yields a statistically significant enhancement in detecting multi-step complex attacks.

\subsubsection{Explainability comparison}

\begin{figure*}[!h]
  \centering
  \subfigure[malicious sample division 78:73:319 ]{
  \includegraphics[width=0.31\linewidth]{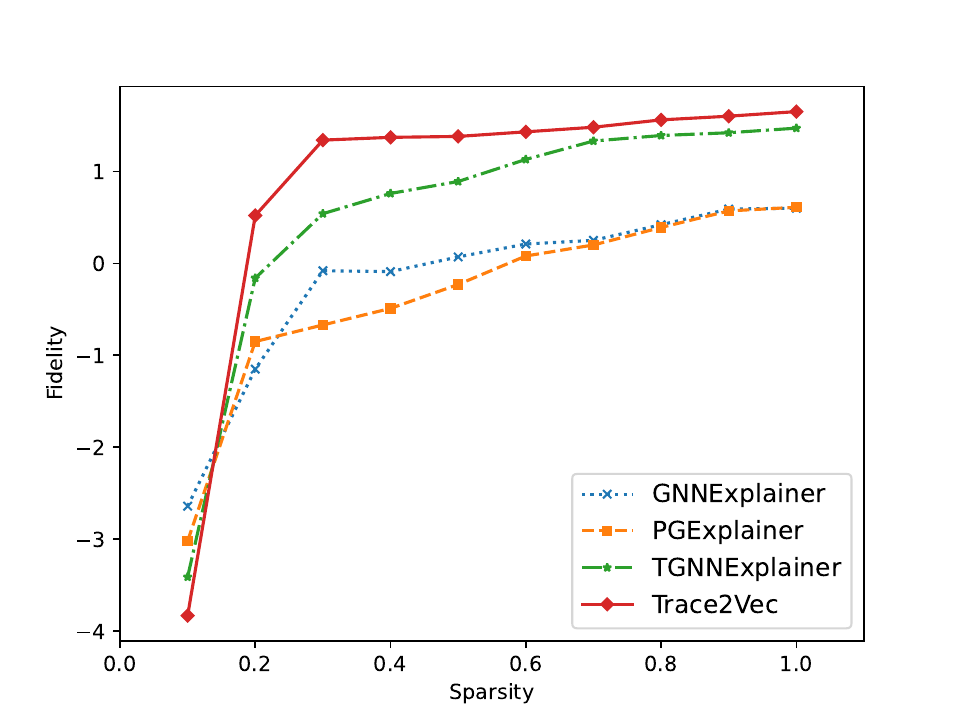}
  }
  \subfigure[malicious sample division 84:182:204]{
  \includegraphics[width=0.31\linewidth]{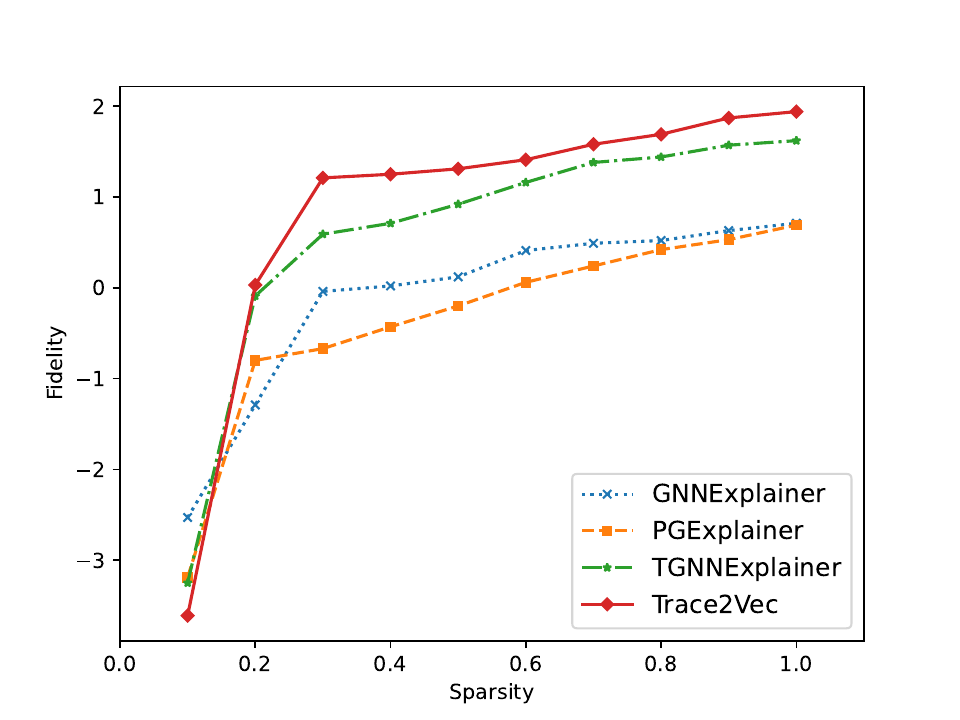}
  }
  \subfigure[malicious sample division 128:168:174]{
  \includegraphics[width=0.31\linewidth]{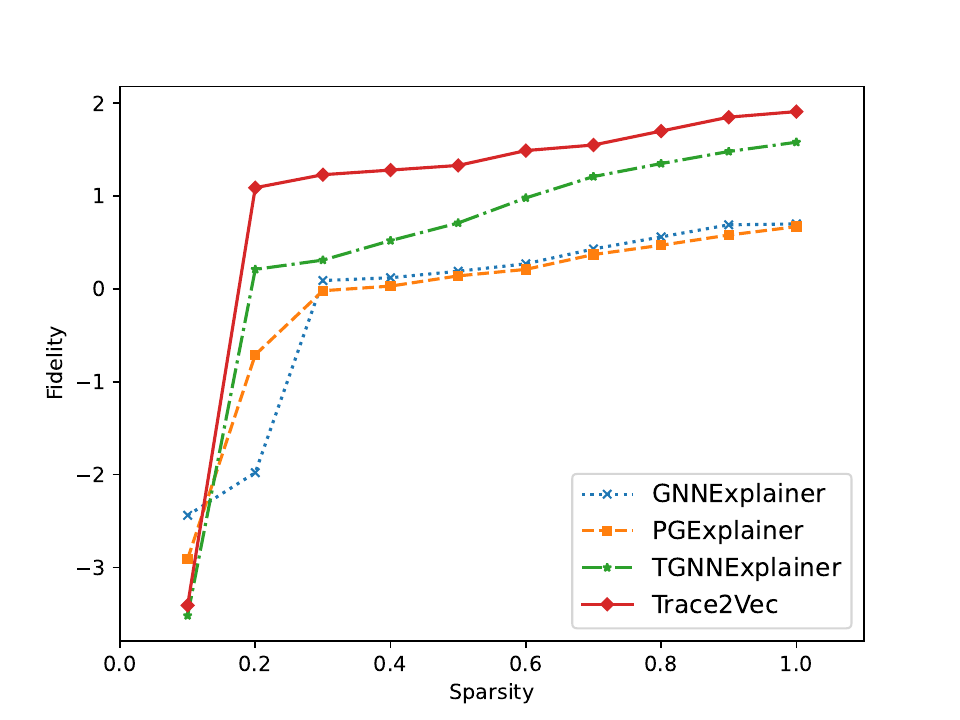}
  }
  \subfigure[malicious sample division 186:152:132]{
  \includegraphics[width=0.31\linewidth]{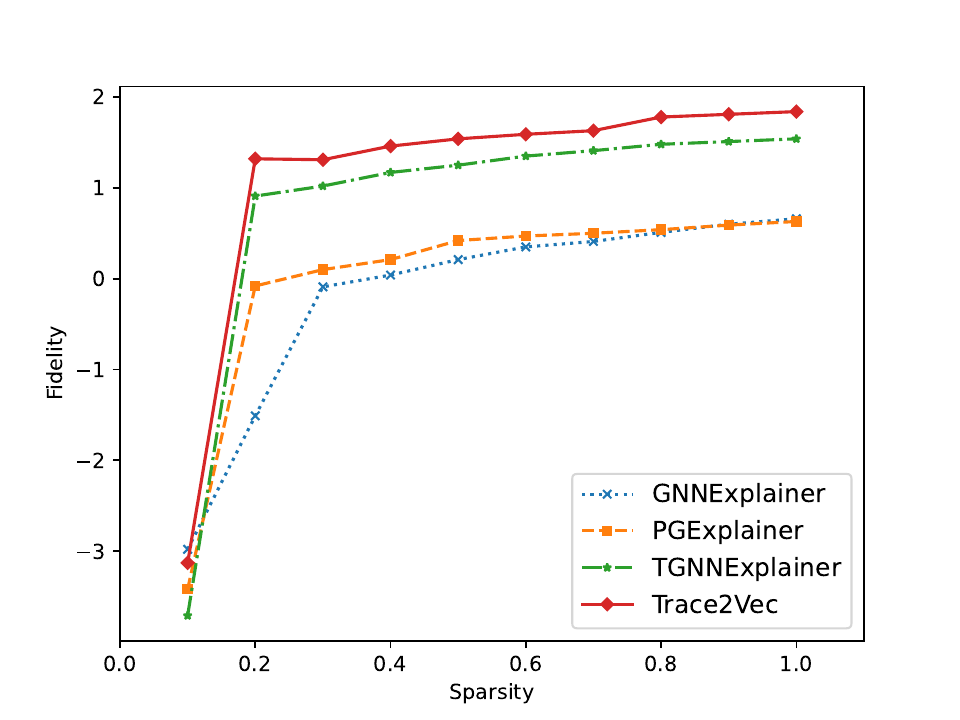}
  }
  \subfigure[malicious sample division 234:132:104]{
  \includegraphics[width=0.31\linewidth]{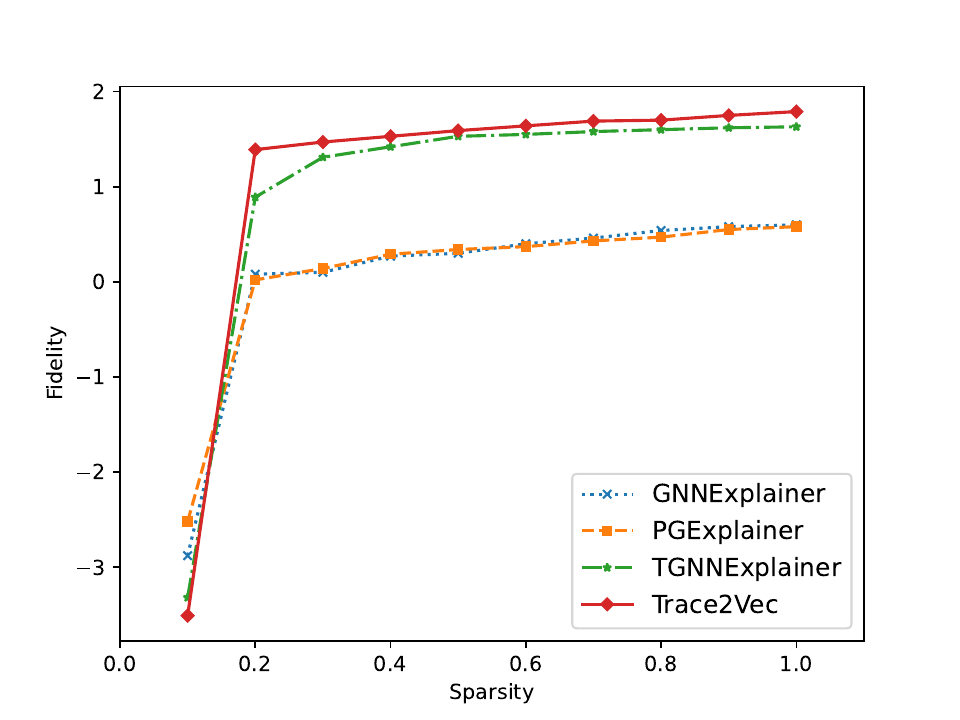}
  }
  \caption{Fidelity-sparsity curves of different divisions of malicious samples on the CERT dataset. X-Axis: the sparsity. Y-Axis: the fidelity. \fullname outperforms the existing competitors.}
  \label{fig:explain}
\end{figure*}

In addition to detection accuracy, a key problem in CMA detection is to ensure the explainability of the detection results, so that trustworthy decisions can be made. 
To this end, Table~\ref{tab:explain} and Table~\ref{tab:explain_atlas} summarize the fidelity and AUFSC results, and Fig.~\ref{fig:explain} depicts the fidelity-sparsity curves. 

\begin{table*}[htbp]
\centering
\caption{Explanation results with different divisions of malicious samples on the CERT dataset.  \fullname outperforms the existing competitors.}
\resizebox{1.0\linewidth}{!}{  
    \begin{tabular}{ccccccccccc}
    \hline
    \multicolumn{3}{c}{\textbf{Malicious samples division}} &  \multicolumn{8}{c}{\textbf{Method}}     \\ \hline
    \multirow{2}{*}{\textbf{\#Train}} & \multirow{2}{*}{\textbf{\#Val}}  & \multirow{2}{*}{\textbf{\#Test}}  & \multicolumn{2}{c}{\textbf{GNNExplainer}} & \multicolumn{2}{c}{\textbf{PGExplainer}} & \multicolumn{2}{c}{\textbf{TGNNExplainer}} & \multicolumn{2}{c}{\bf \fullname} \\ \cline{4-11}
      &    &    & \textbf{Fidelity} & \textbf{AUFSC}   & \textbf{Fidelity} & \textbf{AUFSC}   & \textbf{Fidelity} & \textbf{AUFSC}   & \textbf{Fidelity} & \textbf{AUFSC}   \\ \hline
    78          & 73        & 319        & 0.5989  & 0.5397 & 0.6131  & 0.5187 & 1.4721  & 1.0972 & {\bf 1.6527}  & {\bf 1.3824} \\
    84          & 182       & 204        & 0.7128  & 0.6013 & 0.6891  & 0.5619 & 1.6162  & 1.3627 & {\bf 1.9417}  & {\bf 1.6136} \\
    128          & 168       & 174       & 0.7026 & 0.6138  & 0.6719 & 0.5471  & 1.5839 & 1.3484  & {\bf 1.9120}  & {\bf 1.5915} \\
    186          & 152       & 132       & 0.6571  & 0.5491 & 0.6277  & 0.5311 & 1.5417  & 1.3116 & {\bf 1.8418}  & {\bf 1.5399} \\
    234         & 132       & 104        & 0.5981  & 0.5018 & 0.5831  & 0.5086 & 1.6312  & 1.3189 & {\bf 1.7851}  & {\bf 1.4817} \\ \hline
    \end{tabular}
}
\label{tab:explain}
\end{table*}

\begin{table*}[htbp]
\centering
\caption{Explanation results with different divisions of malicious samples on the ATLAS dataset.  \fullname outperforms the existing competitors.}
\resizebox{1.0\linewidth}{!}{  
    \begin{tabular}{cccccccccccc}
    \hline
    & \multicolumn{3}{c}{\bf Malicious samples division} &  \multicolumn{8}{c}{\bf Method}     \\ \hline
    & \multirow{2}{*}{\bf \#Train} & \multirow{2}{*}{\bf \#Val}  & \multirow{2}{*}{\bf \#Test}  & \multicolumn{2}{c}{\bf GNNExplainer} & 
                               \multicolumn{2}{c}{\bf PGExplainer} & \multicolumn{2}{c}{\bf TGNNExplainer} & \multicolumn{2}{c}{\bf \fullname} \\ \cline{5-12}
                            &            &           &      & \textbf{Fidelity} & \textbf{AUFSC}  & \textbf{Fidelity} & \textbf{AUFSC} & \textbf{Fidelity} & \textbf{AUFSC} & \textbf{Fidelity}      & \textbf{AUFSC}   \\ \hline
 \multirow{3}{*}{M2}        &8923          & 3569        & 5354  & 0.6787  & 0.6369 & 0.7101  & 0.6009 & 1.3194  & 1.1981 & {\bf 1.5987}  & {\bf 1.2985} \\
                            &10708          & 3569       & 3569  & 0.6796  & 0.6400 & 0.7299  & 0.6028 & 1.3198  & 1.1994 & {\bf 1.6491}  & {\bf 1.3219}\\ 
                            &12494         & 2676       & 2676  & 0.6910  & 0.6718 & 0.7372 & 0.6371  & 1.3346 & 1.3589  & {\bf 1.7946}  & {\bf 1.4431} \\ \hline
 \multirow{3}{*}{M5}        &8301         & 2825       & 5907  & 0.6619  & 0.6192 & 0.6922  & 0.5879 & 1.2804  & 1.0877 & {\bf 1.6081}  & {\bf 1.2986} \\
                            &10219         & 3407       & 3407  & 0.6824  & 0.6281 & 0.7171  & 0.5961 & 1.3051  & 1.1182 & {\bf 1.6318}  & {\bf 1.4153} \\
                            &11923         & 2555       & 2555  & 0.7139  & 0.6463 & 0.7419  & 0.6095 & 1.3319  & 1.1291 & {\bf 1.6571}  & {\bf 1.4486} \\ \hline
    \end{tabular}
}
\label{tab:explain_atlas}
\end{table*}

We can observe that \fullname still outperforms all the competitors.
Specifically, on the CERT dataset, when the negative sample ratio is 84:182:204, \fullname achieves the highest fidelity of 1.9417 and the best AUFSC of 1.6136, outperforming the best competitor by 20.1\% and 18.4\%, respectively. Similar improvements are also observed on the ATLAS dataset.
Compared to existing competitors, one of our key difference is to employ the detection module to output the relevance scores. This means that our detection module also helps to improve the explainability by increasing the relevance scores of anomalous behaviors, even in scenarios where malicious samples are extremely rare.
Furthermore, upon inspecting the fidelity-sparsity curves, while at a sparsity level of 0.1, \fullname's fidelity score is slightly lower than that of the other three methods. However, starting from a sparsity value of 0.2, \fullname consistently exhibits higher fidelity values compared to the three baseline methods. This suggests that \fullname efficiently and effectively uncovers associated subsets even at low sparsity levels.
Therefore, comparing with state-of-the-art baseline methods, \fullname demonstrates better explainability in identifying key steps associated with malicious behavior. 

\subsection{Case Study}
To address RQ3 and evaluate the accuracy of \fullname in attack detection and explainability over a long time span, we conducted a simulated augmentation of the CERT dataset by extending the time interval between anomalous behaviors.

\begin{figure*}[!h]
    \centering
    \includegraphics[width=1.0\linewidth, height=0.4\linewidth]{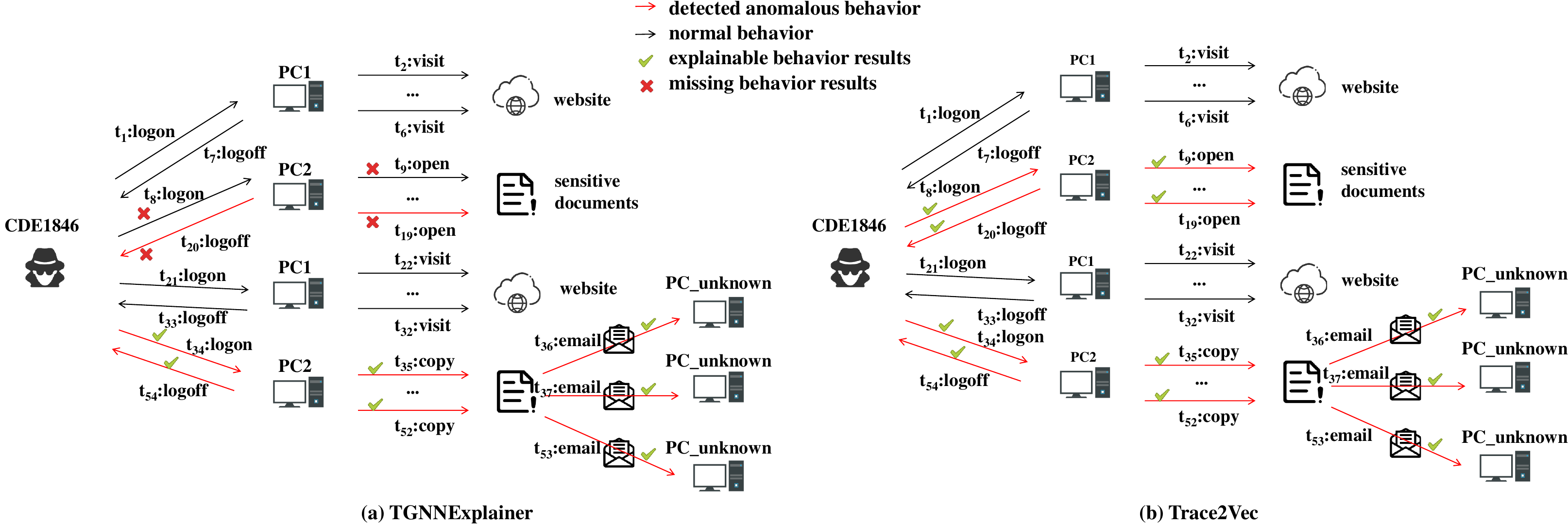}
    \caption{Case study: detection ability of \fullname over long time intervals.}
    \label{fig:casestudy}
\end{figure*}

Taking an attack scenario in this dataset as an instance, which involves an employee with ID CDE1846 logging into another employee's host, searching and copying sensitive files, and sending them to a personal email. This scenario comprises 134 attack samples from February 21, 2011, to April 25, 2011. To capture the characteristic of a long time span, we artificially increased the number of actions performed by the employee CDE1846, simulating normal activities such as email accessing and web browsing. We extended the time span from January 1, 2011, to June 1, 2011. Based on this augmentation, we conducted comparative experiments using \fullname and the best competitor TGNNExplainer. The results are depicted in Fig.\ref{fig:casestudy}, where the red lines represent anomalous behaviors detected by the corresponding methods.

From Fig.\ref{fig:casestudy}(a), it is evident that TGNNExplainer missed the anomalous behaviors at $t_8$ and $t_9$. Furthermore, when tracing the attack behavior at $t_{53}$, it omitted the relevant attack behaviors from $t_8$ to $t_{20}$. This limitation arises because TGNNExplainer struggles to maintain the number of neighbors for each node within a certain threshold. When this threshold is exceeded, it discards distant temporal events, rendering it unable to provide accurate explanations for attacks over long time spans.

From Fig.\ref{fig:casestudy}(b), it can be seen that \fullname identified all the anomalous behaviors, even with a significant time gap between the anomalous behaviors at $t_9$ and $t_{53}$. 
That is, the employee CDE1846 logged into his/her own PC at time $t_1$, browsed web pages normally from $t_2$ to $t_6$, and logged out at $t_7$. Then, at $t_8$, CDE1846 logged into another employee's PC, accessed sensitive files from $t_9$ to $t_{19}$, and logged out at $t_{20}$. Subsequently, from $t_{21}$ to $t_{33}$, CDE1846 browsed web pages normally on his/her own PC. After some time, at $t_{34}$, CDE1846 logged into the other employee's PC again, copied multiple sensitive files, and sent them to an unknown email address. Finally, CDE1846 logged out at $t_{54}$. 
The red-lined portions of the graph represent anomalous behaviors detected by the \fullname model. 
In this scenario, the detection model successfully identified all anomalous behaviors. The checkmarks indicate that \fullname provided explainable traceability results for the attack behavior at $t_{53}$. 


\section{Conclusion}
In this paper, we propose \fullname, an explainable method based on graph neural networks for detecting complex multi-step attacks. \fullname first arguments rare attack samples, and 
applies a continuous-time dynamic heterogeneous graph-based model to detect CMAs. To improve the explainability, \fullname further incorporates the Monte Carlo tree search algorithm, prioritizing events with significant contributions to the attack. 
The experimental evaluation demonstrates the superiority of \fullname over the state-of-the-art approaches. This research introduces a promising avenue for explainable detection, offering robust support to address complex cybersecurity challenges.
As for future work, since \fullname is a post hoc explainable method, we believe that directly embedding an interpretable module into the detection model, i.e., training the explanatory model simultaneously with the detection classifier, to achieve an intrinsic interpretable model will be a key direction.

\hide{
\section*{Acknowledgments}
This was was supported in part by......
}
\bibliographystyle{unsrt}  
\bibliography{references}

\begin{thebibliography}{10}

\bibitem{navarro2018systematic}
Julio Navarro, Aline Deruyver, and Pierre Parrend.
\newblock A systematic survey on multi-step attack detection.
\newblock {\em Computers \& Security}, 76:214--249, 2018.

\bibitem{chen2017automated}
Qian Chen and Robert~A Bridges.
\newblock Automated behavioral analysis of malware: A case study of wannacry ransomware.
\newblock In {\em 2017 16th IEEE International Conference on machine learning and applications (ICMLA)}, pages 454--460. IEEE, 2017.

\bibitem{martinez2021software}
Jeferson Mart{\'\i}nez and Javier~M Dur{\'a}n.
\newblock Software supply chain attacks, a threat to global cybersecurity: Solarwinds’ case study.
\newblock {\em International Journal of Safety and Security Engineering}, 11(5):537--545, 2021.

\bibitem{krombholz2015advanced}
Katharina Krombholz, Heidelinde Hobel, Markus Huber, and Edgar Weippl.
\newblock Advanced social engineering attacks.
\newblock {\em Journal of Information Security and applications}, 22:113--122, 2015.

\bibitem{Hossain2017SLEUTH}
Md~Nahid Hossain, Sadegh~M Milajerdi, Junao Wang, Birhanu Eshete, Rigel Gjomemo, R~Sekar, Scott Stoller, and VN~Venkatakrishnan.
\newblock $\{$SLEUTH$\}$: Real-time attack scenario reconstruction from $\{$COTS$\}$ audit data.
\newblock In {\em 26th USENIX Security Symposium (USENIX Security 17)}, pages 487--504, 2017.

\bibitem{Milajerdi2019POIROT}
Sadegh~M Milajerdi, Birhanu Eshete, Rigel Gjomemo, and VN~Venkatakrishnan.
\newblock Poirot: Aligning attack behavior with kernel audit records for cyber threat hunting.
\newblock In {\em Proceedings of the 2019 ACM SIGSAC conference on computer and communications security}, pages 1795--1812, 2019.

\bibitem{Rossi2020TGN}
Emanuele Rossi, Ben Chamberlain, Fabrizio Frasca, Davide Eynard, Federico Monti, and Michael Bronstein.
\newblock Temporal graph networks for deep learning on dynamic graphs.
\newblock {\em arXiv preprint arXiv:2006.10637}, 2020.

\bibitem{Han2020Unicorn}
Xueyuan Han, Thomas Pasquier, Adam Bates, James Mickens, and Margo Seltzer.
\newblock Unicorn: Runtime provenance-based detector for advanced persistent threats.
\newblock {\em arXiv preprint arXiv:2001.01525}, 2020.

\bibitem{Abdulellah2021ATLAS}
Abdulellah Alsaheel, Yuhong Nan, Shiqing Ma, Le~Yu, Gregory Walkup, Z~Berkay Celik, Xiangyu Zhang, and Dongyan Xu.
\newblock Atlas: A sequence-based learning approach for attack investigation.
\newblock In {\em 30th USENIX security symposium (USENIX security 21)}, pages 3005--3022, 2021.

\bibitem{Zhu2021APTSHIELD}
Tiantian Zhu, Jinkai Yu, Chunlin Xiong, Wenrui Cheng, Qixuan Yuan, Jie Ying, Tieming Chen, Jiabo Zhang, Mingqi Lv, Yan Chen, et~al.
\newblock Aptshield: A stable, efficient and real-time apt detection system for linux hosts.
\newblock {\em IEEE Transactions on Dependable and Secure Computing}, 2023.

\bibitem{Yang2022RShield}
Weiyong Yang, Peng Gao, Hao Huang, Xingshen Wei, Wei Liu, Shishun Zhu, and Wang Luo.
\newblock Rshield: A refined shield for complex multi-step attack detection based on temporal graph network.
\newblock In {\em International Conference on Database Systems for Advanced Applications}, pages 468--480. Springer, 2022.

\bibitem{Ding2023AIRTAG}
Hailun Ding, Juan Zhai, Yuhong Nan, and Shiqing Ma.
\newblock Airtag: Towards automated attack investigation by unsupervised learning with log texts.
\newblock In {\em 32nd USENIX Security Symposium (USENIX Security 23)}, pages 373--390, 2023.

\bibitem{Wei2021Deephunter}
Renzheng Wei, Lijun Cai, Lixin Zhao, Aimin Yu, and Dan Meng.
\newblock Deephunter: A graph neural network based approach for robust cyber threat hunting.
\newblock In {\em Security and Privacy in Communication Networks: 17th EAI International Conference, SecureComm 2021, Virtual Event, September 6--9, 2021, Proceedings, Part I 17}, pages 3--24. Springer, 2021.

\bibitem{Liu2019Log2vec}
Fucheng Liu, Yu~Wen, Dongxue Zhang, Xihe Jiang, Xinyu Xing, and Dan Meng.
\newblock Log2vec: A heterogeneous graph embedding based approach for detecting cyber threats within enterprise.
\newblock In {\em Proceedings of the 2019 ACM SIGSAC conference on computer and communications security}, pages 1777--1794, 2019.

\bibitem{Yang2022Advanced}
Weiyong Yang, Peng Gao, Hao Huang, Xingshen Wei, Haotian Zhang, and Zhihao Qu.
\newblock Advanced persistent threat detection in smart grid clouds using spatiotemporal context-aware graph embedding.
\newblock In {\em GLOBECOM 2022-2022 IEEE Global Communications Conference}, pages 534--540. IEEE, 2022.

\bibitem{Gao2022Detecting}
Peng Gao, Weiyong Yang, Haotian Zhang, Xingshen Wei, Hao Huang, Wang Luo, Zhimin Guo, Yunhe Hao, et~al.
\newblock Detecting unknown threat based on continuous-time dynamic heterogeneous graph network.
\newblock {\em Wireless Communications and Mobile Computing}, 2022, 2022.

\bibitem{Lindauer2020CERT}
Brian Lindauer.
\newblock Insider threat test dataset, 2020.

\bibitem{Yuan2020Xgnn}
Hao Yuan, Jiliang Tang, Xia Hu, and Shuiwang Ji.
\newblock Xgnn: Towards model-level explanations of graph neural networks.
\newblock In {\em Proceedings of the 26th ACM SIGKDD International Conference on Knowledge Discovery \& Data Mining}, pages 430--438, 2020.

\bibitem{Ying2019Gnnexplainer}
Zhitao Ying, Dylan Bourgeois, Jiaxuan You, Marinka Zitnik, and Jure Leskovec.
\newblock Gnnexplainer: Generating explanations for graph neural networks.
\newblock {\em Advances in neural information processing systems}, 32, 2019.

\bibitem{Luo2020Parameterized}
Dongsheng Luo, Wei Cheng, Dongkuan Xu, Wenchao Yu, Bo~Zong, Haifeng Chen, and Xiang Zhang.
\newblock Parameterized explainer for graph neural network.
\newblock {\em Advances in neural information processing systems}, 33:19620--19631, 2020.

\bibitem{Shan2021Reinforcement}
Caihua Shan, Yifei Shen, Yao Zhang, Xiang Li, and Dongsheng Li.
\newblock Reinforcement learning enhanced explainer for graph neural networks.
\newblock {\em Advances in Neural Information Processing Systems}, 34:22523--22533, 2021.

\bibitem{Vu2020Pgm-explainer}
Minh Vu and My~T Thai.
\newblock Pgm-explainer: Probabilistic graphical model explanations for graph neural networks.
\newblock {\em Advances in neural information processing systems}, 33:12225--12235, 2020.

\bibitem{Yuan2021OnExplainability}
Hao Yuan, Haiyang Yu, Jie Wang, Kang Li, and Shuiwang Ji.
\newblock On explainability of graph neural networks via subgraph explorations.
\newblock In {\em International conference on machine learning}, pages 12241--12252. PMLR, 2021.

\bibitem{Wang2020Causal}
Xiang Wang, Yingxin Wu, An~Zhang, Xiangnan He, and Tat seng Chua.
\newblock Causal screening to interpret graph neural networks, 2021.

\bibitem{He2022explainer}
Wenchong He, Minh~N Vu, Zhe Jiang, and My~T Thai.
\newblock An explainer for temporal graph neural networks.
\newblock In {\em GLOBECOM 2022-2022 IEEE Global Communications Conference}, pages 6384--6389. IEEE, 2022.

\bibitem{Xia2023Explaining}
Wenwen Xia, Mincai Lai, Caihua Shan, Yao Zhang, Xinnan Dai, Xiang Li, and Dongsheng Li.
\newblock Explaining temporal graph models through an explorer-navigator framework.
\newblock In {\em The Eleventh International Conference on Learning Representations}, 2022.

\bibitem{Gao2023Deep}
Peng Gao, Haotian Zhang, Ming Wang, Weiyong Yang, Xinshen Wei, Zhuo Lv, and Zengzhou Ma.
\newblock Deep temporal graph infomax for imbalanced insider threat detection.
\newblock {\em Journal of Computer Information Systems}, pages 1--11, 2023.

\bibitem{zhu2018chainsmith}
Ziyun Zhu and Tudor Dumitras.
\newblock Chainsmith: Automatically learning the semantics of malicious campaigns by mining threat intelligence reports.
\newblock In {\em 2018 IEEE European symposium on security and privacy (EuroS\&P)}, pages 458--472. IEEE, 2018.

\bibitem{Velickovic2019DGI}
Petar Veli{\v{c}}kovi{\'c}, William Fedus, William~L Hamilton, Pietro Li{\`o}, Yoshua Bengio, and R~Devon Hjelm.
\newblock Deep graph infomax.
\newblock {\em arXiv preprint arXiv:1809.10341}, 2018.

\bibitem{Kocsis2006Bandit}
Levente Kocsis and Csaba Szepesv{\'a}ri.
\newblock Bandit based monte-carlo planning.
\newblock In {\em European conference on machine learning}, pages 282--293. Springer, 2006.

\bibitem{yuan2022explainability}
Hao Yuan, Haiyang Yu, Shurui Gui, and Shuiwang Ji.
\newblock Explainability in graph neural networks: A taxonomic survey.
\newblock {\em IEEE transactions on pattern analysis and machine intelligence}, 45(5):5782--5799, 2022.

\bibitem{pope2019explainability}
Phillip~E Pope, Soheil Kolouri, Mohammad Rostami, Charles~E Martin, and Heiko Hoffmann.
\newblock Explainability methods for graph convolutional neural networks.
\newblock In {\em Proceedings of the IEEE/CVF conference on computer vision and pattern recognition}, pages 10772--10781, 2019.

\end{thebibliography}

\end{document}